\documentclass[%
 aip,
 amsmath,amssymb,
 reprint,%
]{revtex4-1}

\usepackage{graphicx}
\usepackage{dcolumn}
\usepackage{bm}

\usepackage[utf8]{inputenc}
\usepackage[T1]{fontenc}
\usepackage{mathptmx}

\def\figone{\begin{figure}
\begin{center}
\includegraphics[width=\columnwidth]{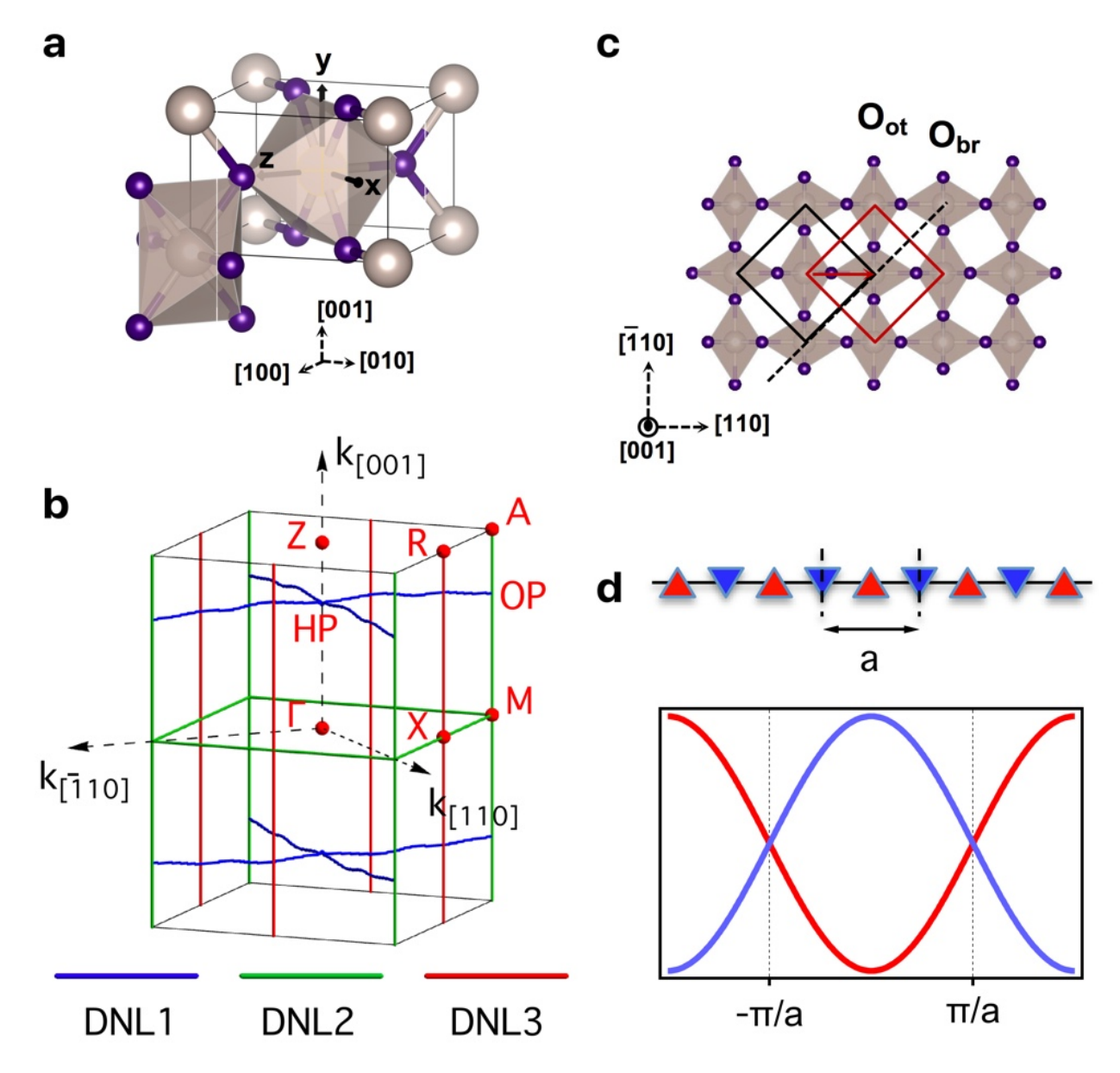}
\caption{\textbf{Dirac nodal lines (DNLs) in RuO$_2$.} 
(a) Crystal structure of RuO$_2$ (Ru: grey, O: purple). Black arrows mark the local coordinate system of the RuO$_6$ octahedron.
(b) BZ of RuO$_2$, summarizing the calculated $k$-space trajectories of the three DNLs.
(c) Crystal structure of RuO$_2$ projected onto the (001) plane, marking two possible choices for the primitive unit cell (black and red squares), as well as the fractional translation (red arrow) and the mirror plane (black dashed line) transforming one into the other. O$_{\text{ot}}$ and O$_{\text{br}}$ label the terminating on-top and bridging oxygen species, respectively. 
(d) Diatomic linear chain with a non-symmorphic glide plane. The translation of half a unit cell in conjunction with mirror reflection produces two bands of opposite parity that form a Dirac crossing at $\pm\pi/a$.
 }
\label{fig1}
\end{center}
\end{figure}}

\def\figtwo{\begin{figure*}[!hbtp]
\begin{center}
\includegraphics[width=\textwidth]{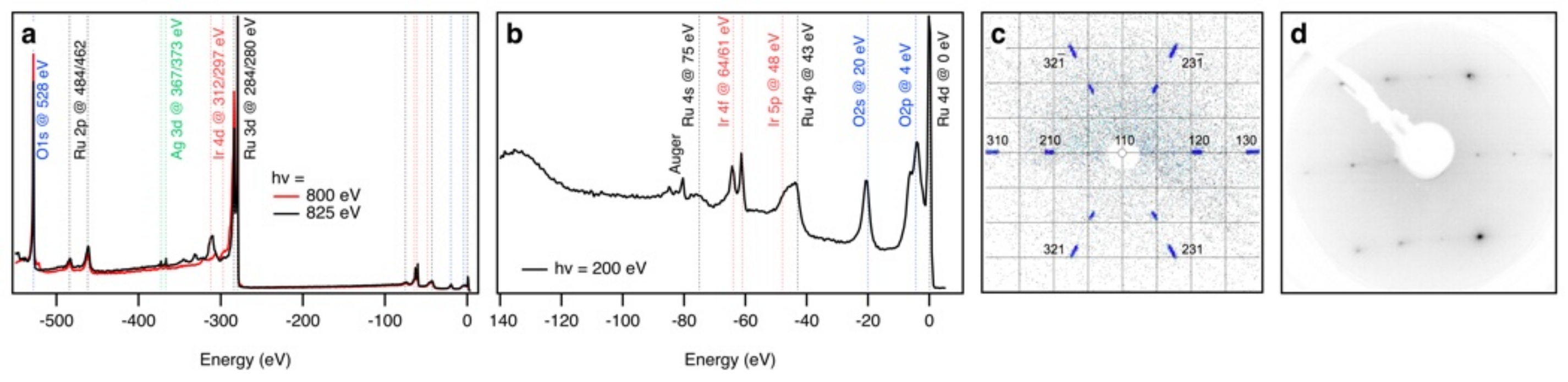}
\caption{\textbf{Surface characterization of RuO$_2$ (110).}
(a) XPS overview and (b) valence band close up of \textit{in situ} cleaved RuO$_2$. Traces of Ag result from residues of the silver epoxy glue used in the pinning/cleaving process, captured by the large beam spot at $\sim 800$~eV photon energy. The Laue diffraction- (c) and LEED pattern (d) confirm the (110) surface orientation.
}
\label{fig2}
\end{center}
\end{figure*}}

\def\figthree{\begin{figure}
\begin{center}
\includegraphics[width=5cm]{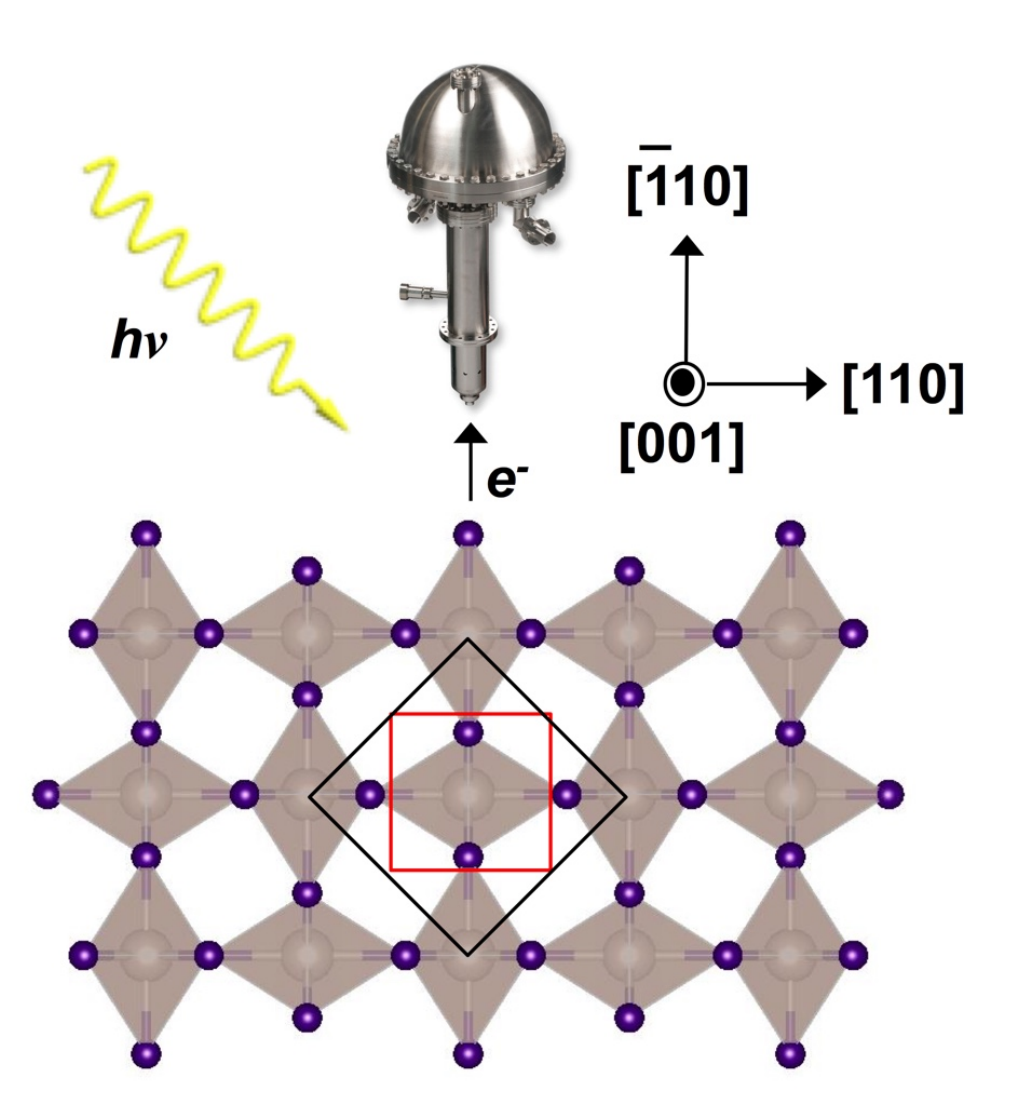}
\caption{\textbf{Experimental ARPES geometry.} The scattering plane is equivalent to the RuO$_2$ (001) crystal plane, with the entrance slit of the electron analyzer oriented along $[110]$, and the surface normal along $[ \overline{1}10]$. The black square is the (001) projection of the primitive unit cell, the red square is the unit cell of the two non-primitive Ru sub-lattices with different RuO$_6$ octahedral orientation.
 }
\label{fig25}
\end{center}
\end{figure}}

\def\figfour{\begin{figure*}[!hbtp]
\begin{center}
\includegraphics[width=\textwidth]{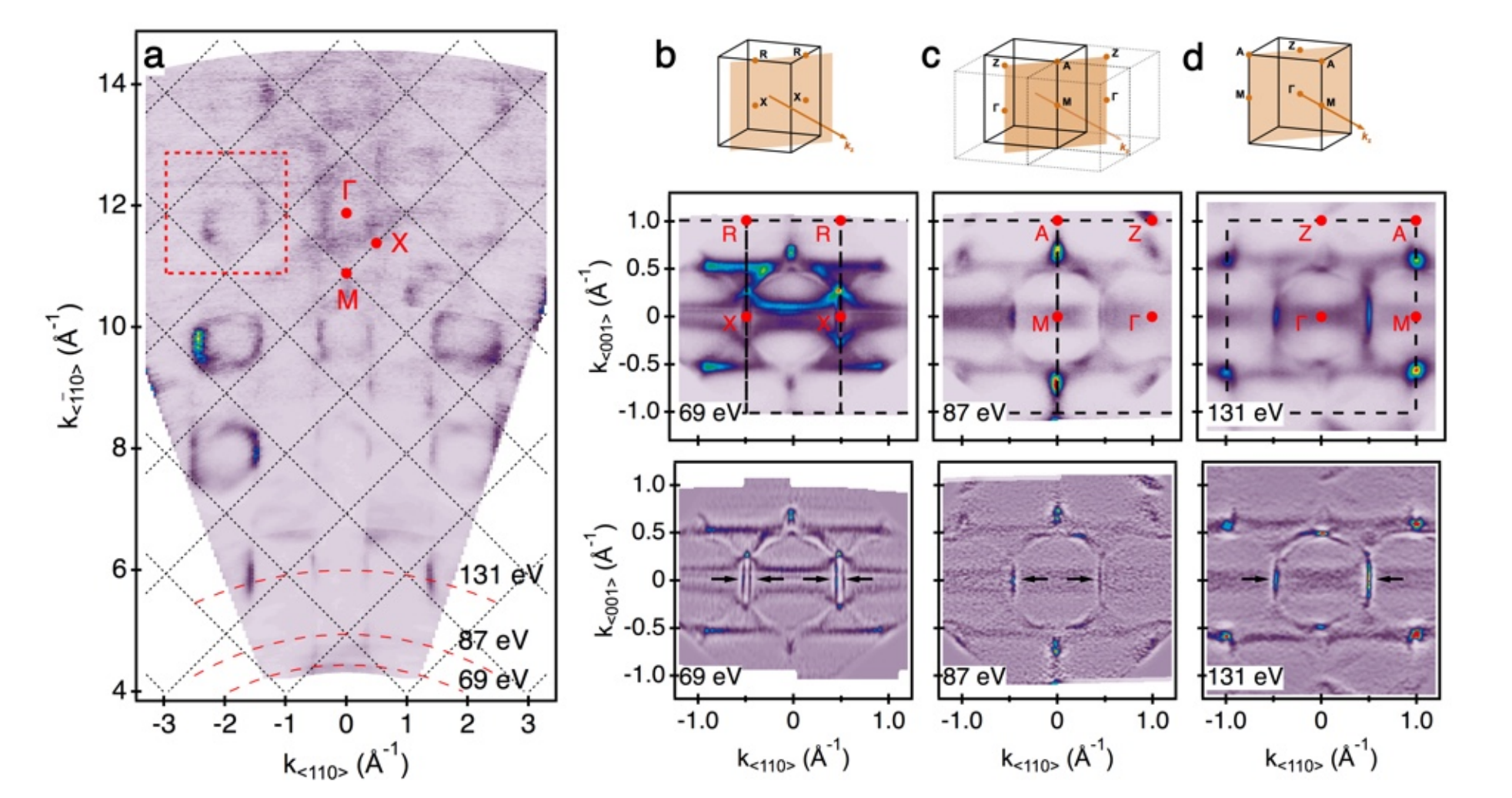}
\caption{
\textbf{(001) Fermi surface along $\Gamma$MX.}  (a) ARPES data compiled over a photon energy range from 60 to 800~eV. Black dotted squares mark the projections of the crystallographic BZ corresponding to the primitive unit cell; the red dashed square marks the projection of the extended BZ corresponding to the unit cell of the Ru sub-lattice (see Fig.~\ref{fig25}). ARPES Fermi surfaces measured with (b) $h\nu=69$~eV, (c) $87$~eV and (d) $131$~eV. The top row indicates the approximate probing plane through the BZ at the respective energy. The bottom row is the curvature of the raw data in the middle row, obtained by a method described in Ref.~\onlinecite{Zhang2011}.
}
\label{fig3}
\end{center}
\end{figure*}}

\def\figfive{\begin{figure*}
\begin{center}
\includegraphics[width=\textwidth]{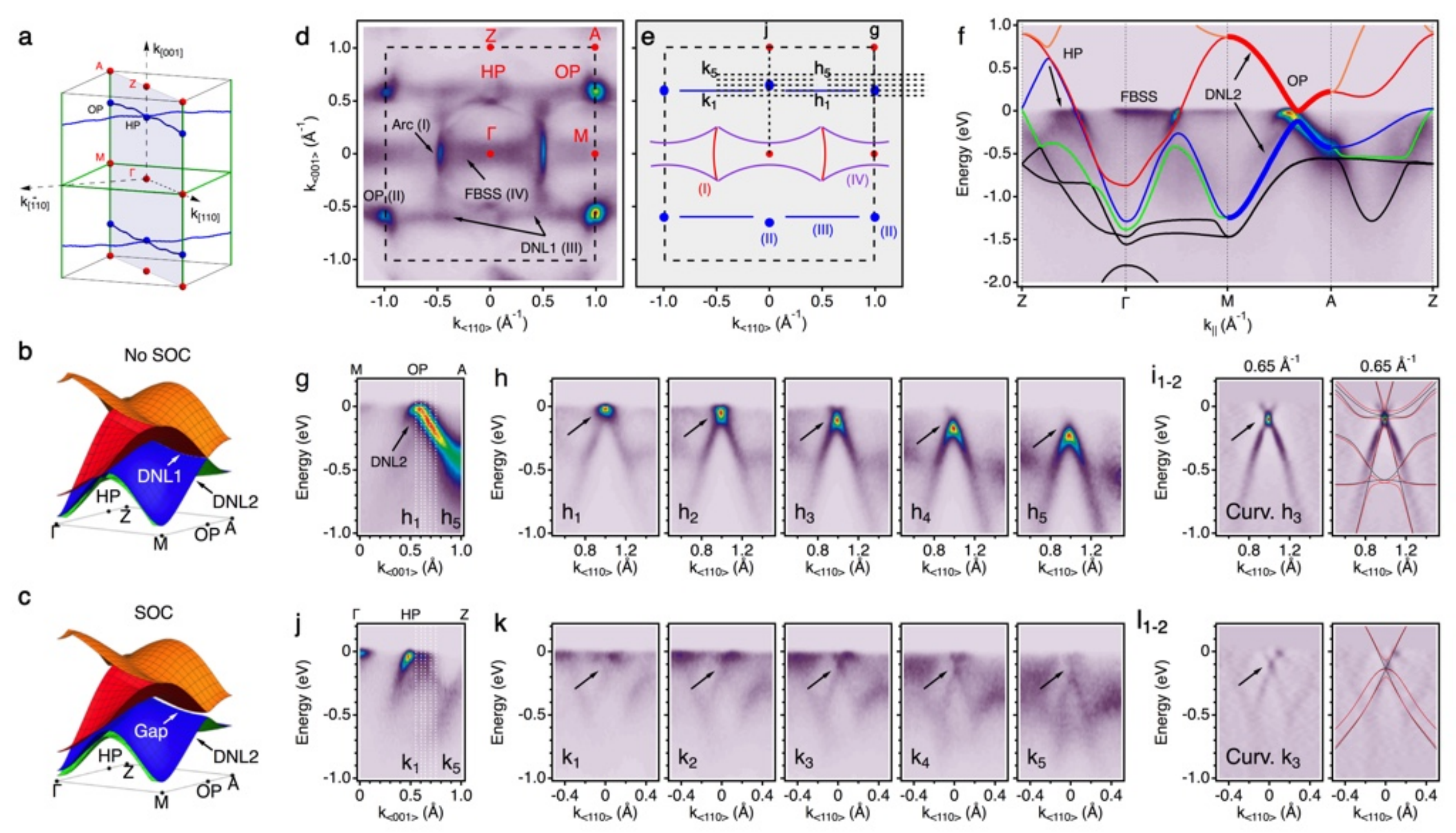}
\caption{\textbf{(110) Fermi surface along $\Gamma$MAZ.} 
(a) BZ of RuO$_2$ highlighting the (110) $\Gamma$MAZ measurement plane, the DNL1 and DNL2, as well as their intersections at points OP and HP.
(b) DFT band-structure model of RuO$_2$ showing DNL1 and DNL2 and their 8-fold degenerate crossing at OP.
(c) SOC gaps DNL1, leaving two sets of 4-fold degenerate bands at OP.
(d) ARPES (110) Fermi surface measured with 131~eV photons, marking high symmetry points (red), BZ boundaries (black dashed) and prominent spectral features (black arrows).
(e) Schematic summary of the Fermi surface in (d). Black dotted lines indicate the position of ARPES cuts in panels (h) and (k).
(f) Comparison between ARPES band structure and DFT+SOC calculations.
(g) Energy dispersion of DNL2 along MA.
(h) Energy dispersion along $k_{\langle 110\rangle}$, showing the evolution of the Dirac crossing close to OP with $k_{\langle 001\rangle}=0.55$ (h$_1$); =0.60 (h$_2$); =0.65 (h$_3$); =0.70 (h$_4$); =0.75 (h$_5$). 
(i) The curvature of the spectrum at $k_{\langle 001\rangle} = 0.65$~\AA$^{-1}$ (i$_1$) \cite{Zhang2011} is compared to LDA-DFT (black) and LDA-DFT+SOC (red) results in (\textbf{i$_2$}).
(j) Energy dispersion of DNL1 along $\Gamma$Z.
(k) Energy dispersion along $k_{\langle 110\rangle}$, showing the evolution of the Dirac crossing close to HP with $k_{\langle 001\rangle}=0.55$ (k$_1$); =0.60 (k$_2$); =0.65 (k$_3$); =0.70 (k$_4$); =0.75 (k$_5$). 
(l) The curvature of the spectrum at $k_{\langle 001\rangle} = 0.65$~\AA$^{-1}$ (l$_1$) \cite{Zhang2011} is compared to shifted (see text) LDA-DFT (black) and LDA-DFT+SOC (red) results in (l$_2$).
}
\label{fig4}
\end{center}
\end{figure*}}

\def\figsix{\begin{figure*}
 \begin{center}
\includegraphics[width=\textwidth]{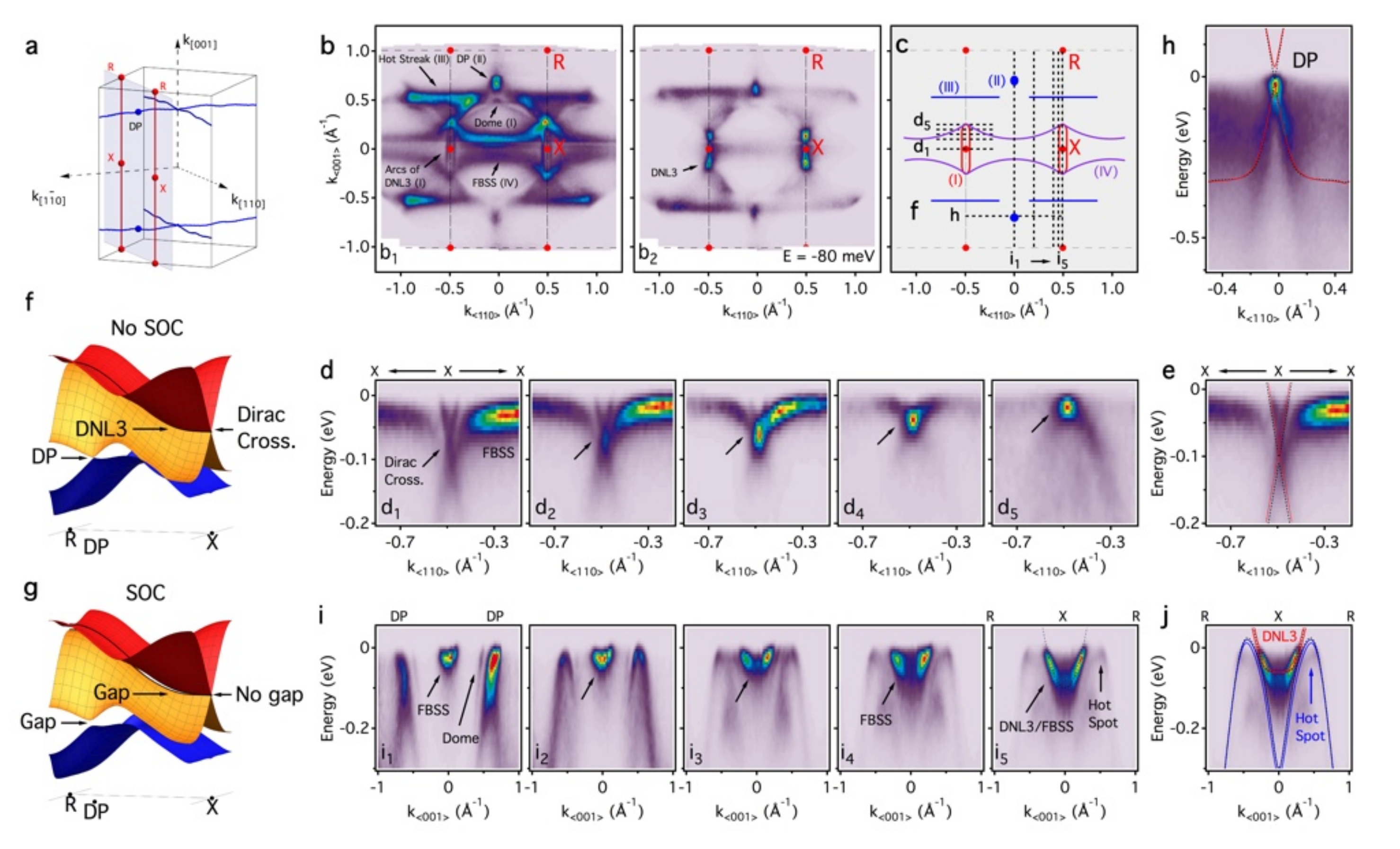}
\caption{\label{fig5} 
\textbf{(110) Fermi surface along XRX.} 
(a) The RuO$_2$ BZ, focusing on the (110) measurement plane containing the X and R high symmetry points, and its intersection with DNL1 at DP.
(b) ARPES Fermi surface (b$_1$) and $E=-80$~meV constant energy cut (b$_2$) measured with 69~eV photons. We mark the high symmetry points (red), the BZ boundaries (black dashed), as well as prominent spectral features (black arrows).
(c) Schematic summary of the Fermi surface in (b). Black dotted lines indicate the position of ARPES cuts in panels (d), (h) and (i).
(d) Energy dispersion along $k_{\langle 110\rangle}$, showing the evolution of the Dirac crossing and the FBSS with $k_{\langle 001\rangle}=0$ (d$_1$); =0.1 (d$_2$); =0.15 (d$_3$); =0.2 (d$_4$); =0.25 (d$_5$). 
(e) Shifted (see text) DFT (black dotted) and DFT+SOC (red solid) calculations, compared to the ARPES data of (d$_1$).
(f) DFT band-structure model of RuO$_2$. 
(g) DFT+SOC band-structure model of RuO$_2$. In comparison to (f), SOC gaps DNL3 and the DP, but the 4-fold band crossing at X is strictly symmetry protected.
(h) ARPES close up of the Dirac crossing at DP, compared to shifted (see text) DFT (black dotted) and DFT+SOC (red solid) calculations.
(i) Energy dispersion along $k_{\langle 001\rangle}$ and the evolution of the FBSS with $k_{\langle 110\rangle}=0$ (i$_1$); =0.2 (i$_2$); =0.4 (i$_3$); =0.45 (i$_4$); =0.495 (i$_5$). Along XR (i$_5$), the FBSS merges with DNL3.
(j) Shifted (see text) DFT (black dotted) and DFT+SOC (red solid) calculations, compared to the ARPES data of (i$_5$). `Hot spot' bands associated with a Fermi surface instability in Ref.~\onlinecite{Berlijn2017} are shown in blue and did not require an energy correction.
}
  \end{center}
\end{figure*}}

\def\figseven{
\begin{figure}
 \begin{center}
\includegraphics[width=\columnwidth]{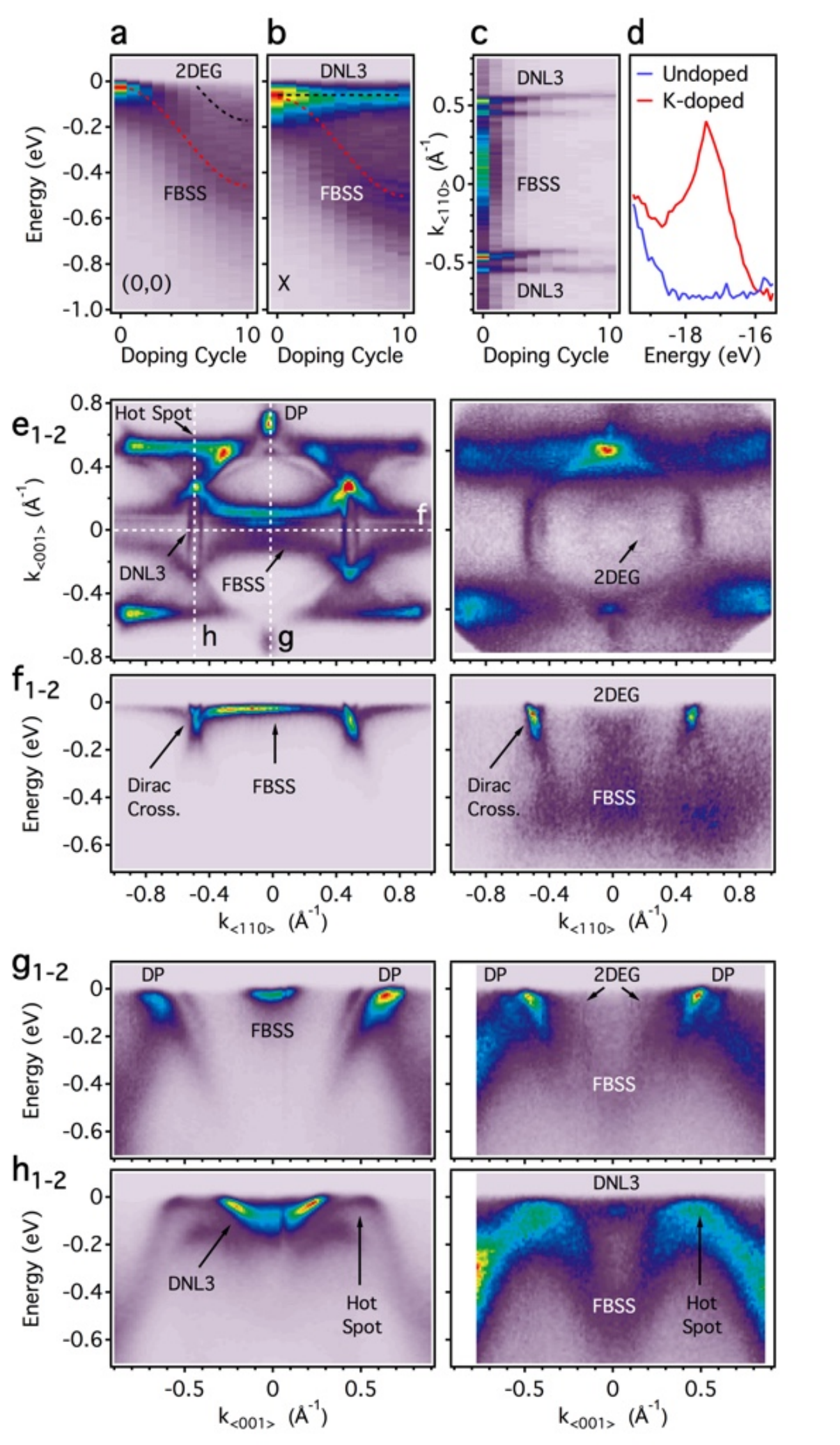}
\caption{\label{fig6} \textbf{Doping evolution of the FBSS.} 
(a) EDC at $(k_{\langle 110\rangle},k_{\langle 001\rangle})=(0,0)$, (b) EDC at $(k_{\langle 110\rangle},k_{\langle 001\rangle})=(0.495,0)$~\AA$^{-1}$ (X), and (c) MDC at $k_{\langle 001\rangle}=0$, all measured \textit{in situ} with 69~eV photons as a function of  potassium deposition. 
(d) K $1s$ core level measurements before (e$_1$) and after deposition (e$_2$). 
(e) Fermi surfaces and (f-h) ARPES band structures along paths indicated in (e$_1$), before (f$_1$-h$_1$) and after (f$_2$-h$_2$) K-deposition.}
  \end{center}
\end{figure}
}

\def\figeight{\begin{figure}[!hbtp]
\begin{center}
\includegraphics[width=\columnwidth]{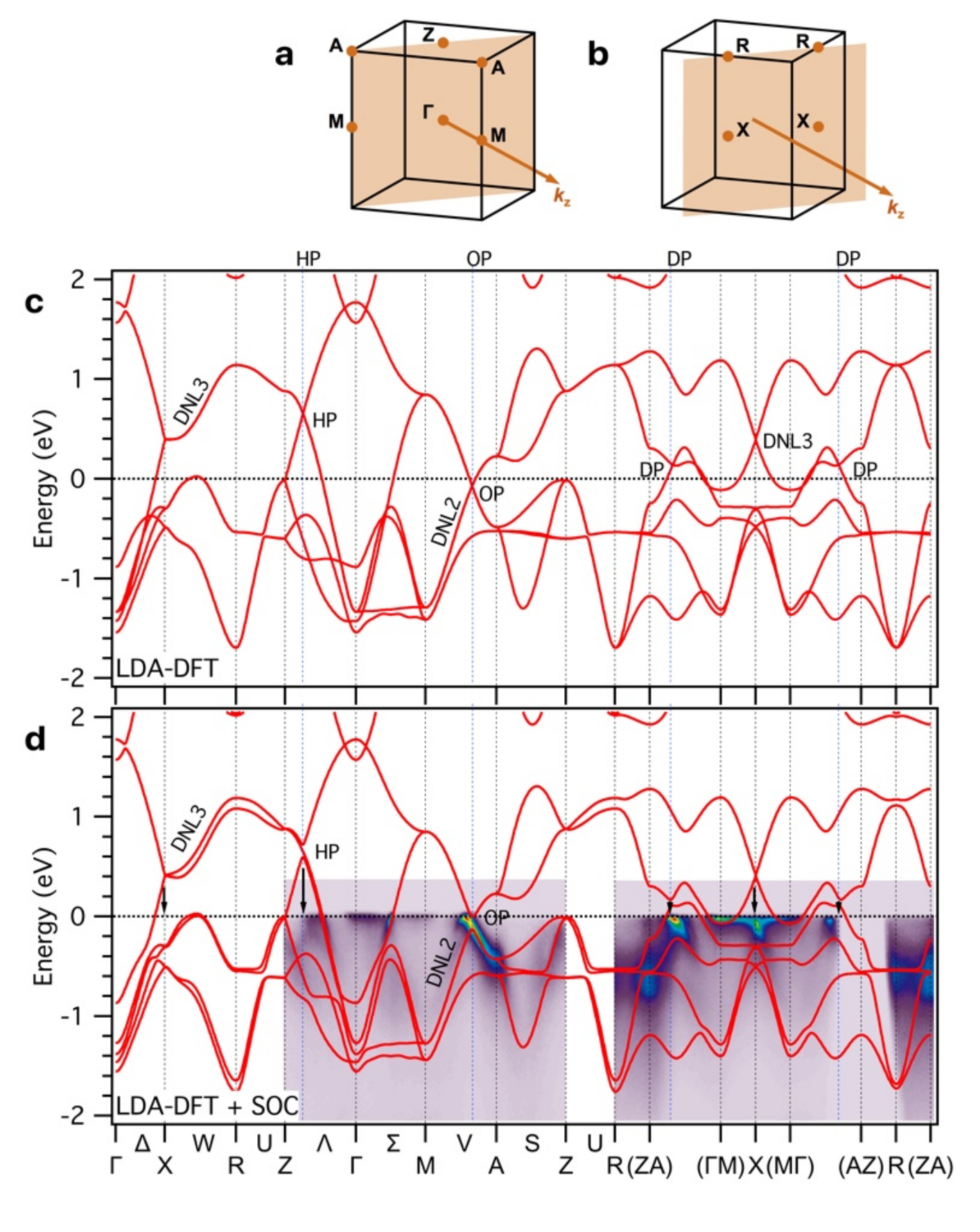}
\caption{\label{fig7} \textbf{ARPES versus DFT.} BZs of RuO$_2$ highlighting the $\Gamma$MAZ (a) and the XRX plane (b), respectively. LDA-DFT calculations along an augmented path are shown without (c) and with SOC (d). Corresponding ARPES intensity maps are plotted for comparison.
}
\end{center}
\end{figure}}

\def\fignine{\begin{figure}[!hbtp]
\begin{center}
\includegraphics[width=\columnwidth]{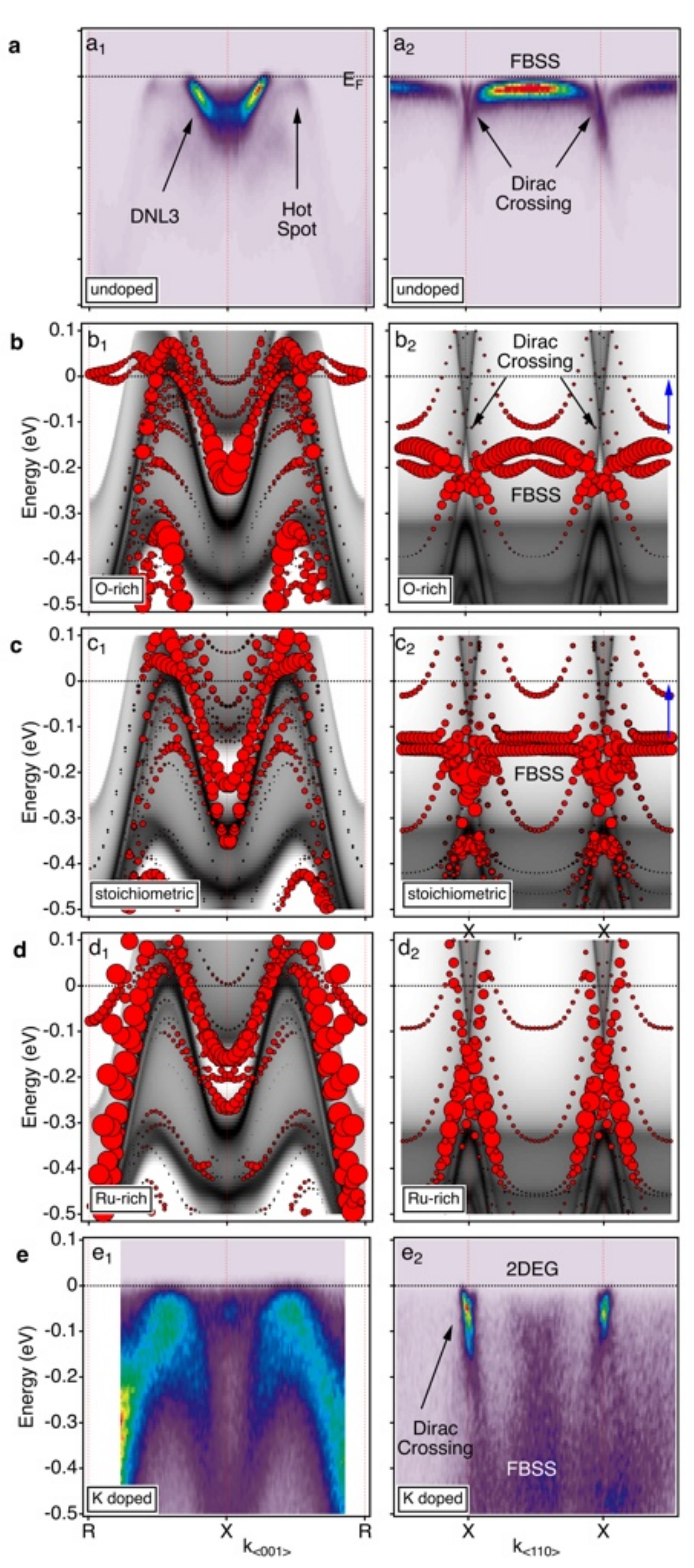}
\caption{\label{fig7_1} \textbf{Surface calculations.} (a) ARPES maps along RXR (a$_1$) and the perpendicular XX direction (a$_2$), outlining DNL3, the `hot spot' states, as well as the FBSS. Corresponding DFT slab calculations of the oxygen rich- (b), the stoichiometric-  (c), and the ruthenium rich RuO$_2$ (110) surface (d). (e) Corresponding ARPES intensity maps of a potassium doped surface.
}
\end{center}
\end{figure}}

\begin{document}

\preprint{AIP/123-QED}

\title[The Dirac nodal line network in non-symmorphic rutile semimetal RuO$_2$]{The Dirac nodal line network in non-symmorphic rutile semimetal RuO$_2$}

\author{Vedran Jovic}
\affiliation{Advanced Light Source, E. O. Lawrence Berkeley National Laboratory, Berkeley, California 94720, USA}
\affiliation{School of Chemical Sciences and Centre for Green Chemical Sciences, The University of Auckland, Auckland 1142, New Zealand}

\author{Roland J. Koch}
\affiliation{Advanced Light Source, E. O. Lawrence Berkeley National Laboratory, Berkeley, California 94720, USA}

\author{Swarup K. Panda}
\affiliation{CPhT, CNRS, Ecole polytechnique, IP Paris, F-91128 Palaiseau, France}

\author{Helmuth Berger}
\affiliation{Institute of Physics, Ecole Polytechnique F\'{e}d\'{e}rale de Lausanne (EPFL), CH-1015 Lausanne, Switzerland}

\author{Philippe Bugnon}
\affiliation{Institute of Physics, Ecole Polytechnique F\'{e}d\'{e}rale de Lausanne (EPFL), CH-1015 Lausanne, Switzerland}

\author{Arnaud Magrez}
\affiliation{Institute of Physics, Ecole Polytechnique F\'{e}d\'{e}rale de Lausanne (EPFL), CH-1015 Lausanne, Switzerland}

\author{Ronny Thomale}
\affiliation{Theoretische Physik I, Universit\"{a}t W\"{u}rzburg, Am Hubland, D-97074 W\"{u}rzburg, Germany}

\author{Kevin E. Smith}
\affiliation{School of Chemical Sciences and Centre for Green Chemical Sciences, The University of Auckland, Auckland 1142, New Zealand}
\affiliation{Department of Physics, Boston University, Boston, Massachusetts 02215, USA}

\author{Silke Biermann}
\affiliation{CPhT, CNRS, Ecole polytechnique, IP Paris, F-91128 Palaiseau, France}

\author{Chris Jozwiak}
\affiliation{Advanced Light Source, E. O. Lawrence Berkeley National Laboratory, Berkeley, California 94720, USA}

\author{Aaron Bostwick}
\affiliation{Advanced Light Source, E. O. Lawrence Berkeley National Laboratory, Berkeley, California 94720, USA}

\author{Eli Rotenberg}
\affiliation{Advanced Light Source, E. O. Lawrence Berkeley National Laboratory, Berkeley, California 94720, USA}

\author{Domenico Di Sante}
\affiliation{Theoretische Physik I, Universit\"{a}t W\"{u}rzburg, Am Hubland, D-97074 W\"{u}rzburg, Germany}

\author{Simon Moser}
\email{simon.moser@physik.uni-wuerzburg.de}
\affiliation{Advanced Light Source, E. O. Lawrence Berkeley National Laboratory, Berkeley, California 94720, USA}
\affiliation{Physikalisches Institut, Universit\"{a}t W\"{u}rzburg, D-97074 W\"{u}rzburg, Germany}

\date{\today}

\begin{abstract}
We employ angle resolved photoemission spectroscopy (ARPES) to investigate the Fermi surface of RuO$_2$. We find a network of two Dirac nodal lines (DNL) as previously predicted in theory, where the valence- and conduction bands touch along continuous lines in momentum space. In addition, we find evidence for a third DNL close to the Fermi level which appears robust despite the presence of significant spin orbit coupling. We demonstrate that the third DNL gives rise to a topologically trivial flat-band surface state (FBSS) at the (110) surface. This FBSS can be tuned by surface doping and presents an interesting playground for the study of surface chemistry and exotic correlation phenomena.
\end{abstract}

\maketitle

\figone
\figtwo
\figthree

\section{Introduction }

Ruthenium dioxide (RuO$_2$) is a functional semi-metal with a wide range of industrial applications, in part stemming from its particular electronic/ionic conduction properties and favorable thermal and chemical stability \cite{Ryan2000,Zhuiykov2011}. RuO$_2$ is corrosion resistant and its diffusion properties are beneficial for pH and dissolved oxygen sensing electrodes, as e.g. employed in water quality monitoring sensors \cite{Zhuiykov2008,Zhuiykov2009,Zhuiykov2009a,Zhuiykov2009b,Zhuiykov2011}. Further, due to particularly high Coulombic efficiencies \cite{Balaya2003} and good mass transport properties, nanoporous RuO$_2$ is a prototypical conversion material in metal oxide lithium-ion battery electrodes with high charge storage capacity (super-capacitors) \cite{Hu2006,Yu2009,Hu2013,Ferris2015}. 

Due to its robustness, RuO$_2$ proves a useful catalyst in a variety of oxidation and dehydrogenation reactions \cite{Over2012,Weaver2013}, such as in the oxidation of carbon monoxide (CO) \cite{Wang2002}, or in the abatement of nitrogen oxides (NO$_x$) from automobile emissions \cite{Wang2011a}. RuO$_2$ further degrades organic molecules such as alcohols \cite{Radjenovic2011}, and dehydrogenates small molecules such as ammonia (NH$_3$) \cite{Cui2010}, which constitutes a useful property  e.g. for wastewater remediation \cite{Li2009,Chiang1995}. Its most important industrial application is the Deacon process, an energy neutral recycling method of hydrochloric acid (HCl) exploiting the exceptional activity of RuO$_2$ for the aniodic evolution of chlorine (Cl$_2$) \cite{Iwanaga2004,Knapp2007,Crihan2008,Zweidinger2008,Seki2010,Teschner2012}, and replacing energy intense conventional recycling methods based on electrolysis \cite{Trasatti2000,Over2012}. RuO$_2$ also has significant potential for energy conversion/storage applications \cite{Over2012,Weaver2013,Trasatti2000}. For instance, RuO$_2$ facilitates the charge separation as a co-catalyst in photocatalytic water splitting \cite{Inoue2009}. The RuO$_2$ (110) surface, in particular, is heavily pursued as a direct catalyst in electrolytic water splitting \cite{Lee2012}, as well as in the electrochemical reduction of carbon dioxide (CO$_2$) at low over-potentials to useful chemical feed-stocks such as methanol (CH$_3$OH) \cite{Karamad2015}.

The surface functionality of RuO$_2$ is rooted in its electronic and magnetic properties -- determined by a complex interplay of lattice-, spin-rotational, and time-reversal symmetries, as well as the competition between Coulomb- and kinetic energies. Consequently, the electronic and magnetic structure of RuO$_2$ has been a matter of longstanding experimental \cite{Ryden1970,Riga1977,Kotz1979,Goel1981,Canart1982,Daniels1984,Cox1986,Kim1997,Kim2004} and theoretical interest \cite{Mattheiss1976,Xu1989,Glassford1993,Yavorsky1996,Yavorsky1996,DeAlmeida2006,ZeJin2010,Mehtougui2012,Torun2013,Ping2015a,Berlijn2017}. Very recently, it was argued that the beneficial catalytic qualities of the RuO$_2$ (110) surface can be directly related to the properties of its Fermi surface. First principle calculations based on density functional theory (DFT) claim magnetic moments on the RuO$_2$ surface to be responsible for low over-potentials in the evolution reaction of ground-state magnetic (triplet) oxygen from nonmagnetic water, resulting in high catalytic efficiencies \cite{Torun2013}. Such local magnetic moments have indeed been confirmed by independent calculations \cite{Ping2015a} as well as neutron \cite{Berlijn2017} and X-ray scattering experiments \cite{Zhu2018}, and are compatible with an antiferromagnetic instability driven by a particularly large density of states at nested `hot spots' in the Fermi surface \cite{Berlijn2017,Ahn2019}. These in turn are believed to be the direct consequence of the non-symmorphic symmetry of the rutile crystal structure, outlining a direct connection between the fundamental symmetry properties and the surface functionality of RuO$_2$ \cite{Torun2013}, and potentially at the heart of novel electronic phenomena such as the crystal Hall effect \cite{Smejkal2019}.

\section{Dirac Nodal Lines in RuO$_2$}

The significance of the non-symmorphic crystal structure of RuO$_2$ was discussed recently within the framework of topology and relativistic Dirac physics \cite{Sun2017}. Symmetry considerations in conjunction with DFT calculations predicted a network of so called Dirac nodal lines (DNL) close to the Fermi level \cite{Burkov2011}, symmetry protected crossings of the conduction and valence band along continuous (one-dimensional) lines in momentum-($k$-) space, with relativistic Fermions close to the intersection. These are in demarcation to the more prominent Dirac \cite{Borisenko2014} and Weyl points \cite{Xu2015a}, where the bands cross in fourfold and twofold degenerate (zero-dimensional) crossing points, respectively. Whereas Weyl points can occur in the absence of any symmetry besides translation, DNLs are typically protected by crystalline symmetries \cite{Fang2015a,Fang2015b,Fang2016}. In particular, they emerge in systems where \textbf{(A)} mirror reflection symmetries \cite{Yang2014,Chiu2014,Bian2016a}, \textbf{(B)} non-symmorphic symmetries (screw axes, glide planes) \cite{Zhao2016a,Li2018,Chiu2018}, or \textbf{(C)} inversion symmetry \cite{Yu2015,Kim2015b,Zhao2016} force a symmetry protected band crossing  (for an overview see Refs.~\onlinecite{Chiu2016,Chan2016,Yu2017,Ekahana2017,Yang2018}). In some cases, this band crossing is accompanied by the formation of flat band surface states (FBSS) spanning in between the surface projection of two DNLs, in analogy to the Fermi arcs spanning in between the surface projections of two Weyl nodes \cite{Burkov2011,Heikkila2011,Weng2015,Chan2016,Wang2017,Huang2017}. Their structure is less universal than in the Weyl case \cite{Kargarian2015,Le2018,Wu2019}, however, where the surface states are immediately protected by topological invariants.

The recent classification of all possible band structures from symmetry principles enabled the identification of a complete set of time-reversal symmetric crystal classes that potentially host DNLs \cite{Bradlyn2017}. Individual studies predicted DNLs explicitly for \textbf{(A)} mirror-symmetric systems such as CaAgX (X = P, As) \cite{Yamakage2016}, Ca$_3$P$_2$ \cite{Xie2015,Chan2016}, YH$_3$ \cite{Kobayashi2017}, TlTaSe$_2$ \cite{Bian2016},  PbTaSe$_2$ \cite{Bian2016a}, and the fcc alkaline earth metals (Ca, Sr, Yb) \cite{Hirayama2016}, for \textbf{(B)} non-symmorphic systems such as X$_3$SiTe$_6$ (X = Ta, Nb) \cite{Li2018}, Ag$_2$S \cite{Huang2017}, ReO$_2$ \cite{Wang2017}, or hyper-honeycomb lattices \cite{Mullen2015}, and for \textbf{(C)} centro-symmetric anti-perovskites such as Cu$_3$PdN \cite{Yu2015}, or three-dimensional graphene networks \cite{Weng2015}. Further, tunable DNLs have been predicted for systems without time reversal symmetry such as hetero-structures made from alternating layers of topological- and magnetic insulators \cite{Phillips2014}. Experimental observations of DNLs, in particular accompanied by a FBSS, remain scarce to this date, but have been reported for \textbf{(A)} mirror-symmetric two-dimensional monolayers of Cu$_2$Si \cite{Feng2016a}; for \textbf{(B)} non-symmorphic materials such as XSiS (X= Zr, Hf) \cite{Schoop2015,Neupane2016,Wang2016,Chen2017} as well as InBi \cite{Ekahana2017}; and for \textbf{(C)} centro-symmetric metal di-borides such as ZrB$_2$ \cite{Zhang2017,Liu2018,Feng2018,Lou2018}. 
 
\figfour

In RuO$_2$ (and its sister compounds IrO$_2$ and OsO$_2$), whose rutile crystal structure and Brillouin zone (BZ) are outlined in Fig.~\ref{fig1}~(a) and (b), symmetry predicts two types of DNLs close to the Fermi level \cite{Sun2017}: \textbf{(A)} First, time reversal- and inversion symmetry in unison with a mirror symmetry protect a band crossing within the (110) and $(\overline{1}10)$ planes. This produces a network of 4-fold degenerate (2 $\times$ spin and 2 $\times$ orbital) and topologically nontrivial DNLs (DNL1), outlined by blue lines in the Brillouin zone (BZ) of Fig.~\ref{fig1}~(b). \textbf{(B)} The second type of DNL in RuO$_2$ is topologically trivial and protected by a non-symmorphic glide mirror symmetry. In a nutshell, the RuO$_2$ lattice is a bipartite composition of two sub-lattices with different RuO$_6$ octahedral orientation. As illustrated in Fig.~\ref{fig1}~(c), these are related by a fractional lattice translation of half a body diagonal (red arrow) and a reflection about the (100) plane (black dashed line) that takes the crystal into itself. In analogy with the di-atomic chain model in Fig.~\ref{fig1}~(d), such a glide plane effectively doubles the unit cell and back-folds the electronic bands. In conjunction with time reversal symmetry it produces fourfold (2 $\times$ spin and 2 $\times$ orbital) degenerate DNL2s along the $k_x=\pi/a$ and $k_y=\pi/a$ boundary planes of the crystallographic BZ \cite{Young2015,Yang2018}. 

The intersections of the DNLs along $\Gamma$Z and MA are 6-fold and 8-fold degenerate, specifying these points as hexatruple (HP) and octuple points (OP), respectively (Fig.\ref{fig1}~b). In the presence of spin orbit coupling (SOC) the DNL1s were predicted to gap out (in particular at OP and HP), resulting in the anti-crossing of the Ru 3$d$ manifold, which is expected to produce large intrinsic spin-Hall conductivities \cite{Sun2017} similar as reported for IrO$_2$ \cite{Fujiwara2013,Das2018}. SOC will also partially gap the DNL2s along the BZ boundary, but additional crystalline symmetries protect the 4-fold band degeneracy along the discrete high symmetry lines XM and MA (green lines in Fig.\ref{fig1}~b) \cite{Sun2017, Burkov2011}.

Building up on our previous rapid communication \cite{Jovic2018}, we support and detail this scenario by state-of-the-art micro focused angle resolved photoemission spectroscopy ($\mu$ARPES) on the (110) surface of 7~\% Ir doped RuO$_2$ single crystals. Beyond the predictions of  DFT \cite{Sun2017}, we find clear signatures of an additional, topologically trivial but unexpected DNL3 of type (B) along the XR direction in the BZ (red lines in Fig.~\ref{fig1} b), producing a continuous Dirac crossing in close proximity to the Fermi level, and remaining surprisingly intact despite considerable SOC. This DNL3 serves as an anchor line for a non-dispersive FBSS, a trivial analogue of the theoretically predicted topological drumhead surface state \cite{Burkov2011,Heikkila2011,Weng2015,Chan2016,Wang2017,Huang2017}. Close to the Fermi level, the density of states of this FBSS diverges in a van Hove singularity like fashion, and possibly gives rise to novel exotic surface phenomena such as unconventional superconductivity \cite{Kopnin2011,Tang2014}, surface magnetism \cite{Magda2014,Chan2016}, long-range Coulomb interaction \cite{Huh2016} or graphene-like Landau levels \cite{Rhim2015}. With its strong response to changes in the electrostatic environment, this FBSS is likely also involved in surface catalytic processes. Finally, we reveal the nested `hot spot' features that are claimed responsible for antiferromagnetic Fermi surface instability scenarios in RuO$_2$ \cite{Berlijn2017,Zhu2018,Ahn2019}, a postulated pillar of its catalytic efficiency \cite{Torun2013}, and of the recently predicted crystal Hall effect \cite{Smejkal2019}.

\section{Crystal growth and characterization}

Ir doped RuO$_2$ single crystals were grown by chemical vapor transport at the Crystal Growth Facility of the EPFL in Lausanne, Switzerland, following a recipe described in Ref.~\onlinecite{Schafer1963}: A powder of 90~mol\% RuO$_2$ and 10~mol\% IrO$_2$ was encapsulated in a quartz ampule (diameter 2~cm, length 15~cm) with TeCl$_4$ used as a transport agent. The growth proceeded with a source temperature of 1060~$^\circ$C and a temperature gradient of 50~$^\circ$C. After about two weeks, millimeter sized single crystals were obtained. X-ray diffraction (XRD, not shown) measurements confirm the rutile crystal structure (Fig.~\ref{fig1}~a), with lattice parameters  $a=4.48$~\AA~and $c=3.105$~\AA~as reported in literature \cite{Cotton1966,Butler1971}. X-ray fluorescence (XRF, not shown) and an analysis of the XPS core level spectra in Fig.~\ref{fig2}~(a) and (b) verify a doping level of 7~\% Ir in RuO$_2$: Ru$_{0.93}$Ir$_{0.07}$O$_2$. This doping slightly raises the Fermi level, but leaves the overall electronic band structure unaffected.

\section{\bf{$\mu$}ARPES experiments}
The $\mu$ARPES experiments were performed at the Microscopic and Electronic Structure Observatory (MAESTRO), beamline 7.0.2 of the Advanced Light Source (ALS) in Berkeley, USA. The ultra high vacuum (UHV) $\mu$ARPES end-station had a base pressure better than $5 \times 10^{-11}$~mbar. RuO$_2$ single crystals were mounted with a ceramic top pin and cleaved \textit{in vacuo}, reproducibly exposing small ($\sim50~\mu$m), clean facets of the oxygen terminated (110) surface \cite{Over2012,Weaver2013}. The orientation was confirmed by Laue diffraction (Fig.~\ref{fig2}~c) and LEED (Fig.~\ref{fig2}~d). 

The synchrotron beam-spot size on the sample was smaller than the domain size on the order of $\lesssim 20~\mu$m for photon energies below 200~eV. The ARPES data were collected with $p$-polarized light, with the polarization vector as well as the analyzer entrance slit in the horizontal scattering plane. The $[110]$ crystal axis was oriented along the analyzer slit, with the $[\overline{1}10]$ surface vector pointing towards the analyzer (Fig.~\ref{fig25}).

The lens optics of our hemispherical Scienta R4000 electron analyzer is equipped with custom-made electrostatic deflectors. These enabled us to collect ARPES spectra over an entire surface BZ without moving the sample, allowing us to retain the X-ray focus on a small sample facet without drift. The total energy and momentum resolution of the experimental setup (beamline \& analyzer) were better than 20~meV and 0.01~\AA$^{-1}$, respectively. Measurements and dosing experiments were carried out below 70~K. Potassium dosing experiments were carried out \textit{in situ} using SAES getters mounted in the $\mu$ARPES chamber, such that dosing could be performed on an optimized sample position without moving the sample.

\section{DFT calculations}

Bulk band structure calculations for RuO$_2$ were performed within density-functional theory (DFT), employing the full potential linearized augmented plane wave (FP-LAPW) method and a non-spin polarized setup as implemented in the Wien2k code \cite{Blaha1990}. The experimental crystal structure and atomic positions were taken from Ref.~\onlinecite{Butler1971}. Exchange and correlation effects were treated using the local density approximation (LDA) with and without including spin-orbit coupling effects. The Brillouin- Zone integration has been performed using a $10\times10\times14$ $k$-mesh. To achieve energy convergence of the eigenvalues, the wave functions in the interstitial region were expanded in plane waves with a cutoff $R_{\text{MT}}k_{\text{max}}=7$, where $R_{\text{MT}}$ denotes the smallest atomic sphere radius and $k_{\text{max}}$ represents the magnitude of the largest $k$-vector in the plane wave expansion. The maximum value of the angular momentum ($l_{\text{max}}$) was taken equal to 10 for the expansion of valence wave functions inside the spheres, while the charge density is Fourier-expanded up to $G_{\text{max}}$ (magnitude of the largest vector) = 12 (a.u.)$^{-1}$. 

In addition to the DFT analysis, we have also derived a low-energy tight-binding (TB) model for the Ru $4d$ states to theoretically analyze our ARPES spectra along various $k$-directions, and to construct the 3D band structure models shown in this work. The hopping amplitudes between the effective $4d$-states and their on-site energies were obtained by constructing the Wannier function for the Ru-$4d$-like bands using the WANNIER90 \cite{Mostofi2014} and WIEN2WANNIER \cite{Kunes2010} codes. The on-site SOC contribution of the Ru $4d$ orbitals to the TB Hamiltonian was modeled by a single parameter $\lambda=120$~meV fitted to the DFT+SOC bandstructure, a value considerably lower than the calculated atomic limit of $\sim167$~meV \cite{Dunn1961,Shanavas2014}.

Surface band structure calculations of the RuO$_2$ (110) surface were performed by means of the Vienna ab initio Simulation Package (VASP) \cite{Kresse1996}. We used a plane wave cutoff of 600~eV on a $4\times 4 \times 1$ Monkhorst-Pack $k$-point mesh, and SOC has been self-consistently included. The slab consisted of 10 octahedral RuO$_2$ layers which account for a thickness of $~32$~\AA. More than $~10$~\AA~of vacuum is included to screen the interaction between the repeated images in the periodic boundary conditions setting.

\figfive
\figsix

\section{(001) Fermi surface along $\Gamma$MX.}\label{sec: Navigating $k$-space}

To identify the high symmetry planes within the complex 3D band structure of RuO$_2$, we employed photon energy ($h\nu$) dependent $\mu$ARPES. Figure~\ref{fig3}~(a) shows the experimental Fermi surface in the (001) plane, compiled from ARPES measurements in the $h\nu$ range 60~eV to 800~eV, and employing an inner potential of $V_0=15$~eV that is consistent with the periodicity of the data. The projection of the crystallographic 3D Brillouin zone (BZ) is marked in black dotted and corresponds to the black primitive unit cell in Fig.~\ref{fig25}. The projection of the extended BZ is marked in red dashed and corresponds to the red non-primitive unit cell of the Ru sub-lattice in Fig.~\ref{fig25}. We find circular spectral contours, enhanced in every other crystallographic BZ, and thus following the non-primitive Ru-sublattice-, rather than the primitive periodicity. This checkerboard signature is caused by the constructive/destructive interference of photoelectrons emitted from the two individual Ru-sublattices, a consequence of the non-symmorphic glide mirror plane of the rutile RuO$_2$ crystal structure \cite{MoserME2016}.

From the periodicity along $[ \overline{1}10]$, we identify photon energies where the ARPES probing sphere cuts the 3D BZ close to the high symmetry planes (e.g. red dashed bows in Fig.~\ref{fig3}~a). At these energies, we employ the custom electrostatic deflectors of our photo-electron analyzer to acquire full $k_{\langle 110\rangle}$-$k_{\langle 001\rangle}$-$E$ ARPES datasets, summarized in Fig.~\ref{fig3}~(b-d). As illustrated at the top of panel (b), the ARPES hemisphere at $h\nu = 69$~eV photon energy cuts the 3D BZ close to a (110) plane that contains both the $X$ and $R$ high symmetry points.

In contrast, the ARPES hemisphere at  (c) $h\nu=87$~eV and (d) $h\nu=131$~eV photon energy probes the 3D BZ close to a (110) plane identical with the $\Gamma$MAZ high symmetry plane, and the normal emission vector identifying with the M- or the $\Gamma$-point, respectively. The raw ARPES Fermi-surfaces shown in the middle row are complemented by their curvatures in the bottom row. These were obtained by a method described in detail in Ref.~\onlinecite{Zhang2011} and trace sharper Fermi surface features as compared to the raw data.

\section{(110) Fermi surface along $\Gamma$MAZ.}

Having identified the high symmetry planes, we now confirm the theoretical predictions of DNLs in RuO$_2$ \cite{Sun2017}. We focus on the $\Gamma$MAZ high symmetry plane, outlined in Fig.~\ref{fig4}~(a). Fig.~\ref{fig4}~(b) is a 3D DFT band structure model, revealing the continuous Dirac crossing of DNL1 and DNL2. In the presence of SOC, the DNL1s are expected to gap, as illustrated in Fig.~\ref{fig4}~(c), the fourfold degeneracy along the DNL2s, however, remain strictly symmetry protected along the XM and MA lines in the BZ (green lines in Fig.~\ref{fig4}~a) \cite{Sun2017, Burkov2011}. 

Our ARPES data confirms this scenario: Figure~\ref{fig4}~(d) shows an ARPES Fermi map measured with $h\nu = 131$~eV photon energy. As discussed in section~\ref{sec: Navigating $k$-space}, the photoemission hemisphere at this energy probes the RuO$_2$ BZ along the $\Gamma$MAZ plane. With the lattice parameters $a=4.49$~\AA~and $c=3.11$~\AA~of RuO$_2$, the (110) surface BZ projection (black dashed line in panel d) is almost quadratic: $a^*\sqrt{2} = 1.98$~\AA$^{-1}$ and $c^* =2.02$~\AA$^{-1}$. We identify four main contributions to the Fermi surface, summarized in the schematics of panel (e), and well reproduced by theory (see e.g. Fig.~4 in Ref.~\onlinecite{Ahn2019}). 

(I) First, we observe arc structures centered half way between $\Gamma$ and M at $k_{\langle110\rangle}=\pm a^*\sqrt{2}/4 =\pm0.495$~\AA$^{-1}$. As we will see later, these represent one branch of the Dirac crossing that forms DNL3. The arcs extend towards the zone center, and form a faint dome. (II) Second, we find intense spectral features labeled OP along MA \cite{Sun2017}. These are the remnant intersection points of DNL1 and DNL2, outlined in panel (a), and identify with the `hot-spots $K_2$' claimed responsible for an antiferromagnetic Pomeranchuk instability in Ref.~\onlinecite{Ahn2019}. They are complimented by weaker features HP along $\Gamma$Z, where the DNL1s cross each other. (III) Third, we find fuzzy horizontal streaks, connecting points OP and HP, and representing the SOC split remnants of DNL1. As we will see later, these features are connected to another set of `hot spots' along XR claimed responsible for an antiferromagnetic spin-density wave instability in Ref.~\onlinecite{Berlijn2017}. (IV) Last, we identify two prominent horizontal arcs spanning in between the arc features (I), the signature of the FBSS. 

The ARPES band structure along the high symmetry path Z$\Gamma$MAZ is shown in Fig.~\ref{fig4}~f, well traced by our DFT+SOC band-structure, and well reproduced by the ARPES intensity calculations in Fig.~3 of Ref.~\onlinecite{Ahn2019}. Following the discussion in Ref.~\onlinecite{Sun2017} (in particular of Fig.~1~(f) therein), we directly identify the ARPES band feature along MA with DNL2. Along M-OP, the orbital character is primarily $d_{xz}$, of even parity with respect to the $(001)$ ARPES scattering plane, and given the odd $p$-polarization vector, of suppressed spectral weight (see Fig.~\ref{fig1}~(a) for definition of the local RuO$_6$ octahedra coordinate system $(x,y,z)$). Along OP-A, the orbital character is mostly $d_{yz}$, and the band thus clearly visible in ARPES \cite{Ahn2019}. Surprisingly, we also find the remnants of the hexatruple band crossing about 0.13~eV below the Fermi level, i.e. about 0.79~eV below the prediction of bulk DFT (shown by the arrow in f). As to be expected, our bulk DFT description also misses the non-dispersive FBSS along $\Gamma$M.

An expanded view of the DNL2 dispersion along MA is shown in Fig.~\ref{fig4}~(g), and highlights the band maximum at OP ($k_{\langle 001 \rangle}\sim 0.55$~\AA$^{-1}$) close to the Fermi level. Perpendicular ARPES band structure cuts (panels h$_{1-5}$), extracted at prominent momenta $k_{\langle 001 \rangle}$ (black dashed lines in \ref{fig4}~e and white lines in \ref{fig4}~g), reveal the appearance and evolution of a Dirac crossing between OP and A, the signature of DNL2. The curvature of panel (h$_3$) \cite{Zhang2011}, displayed in panel (i$_1$), exemplifies this crossing, and shows excellent agreement with the bulk DFT prediction in panel (i$_2$). In an analogous manner, Fig.~\ref{fig4}~(j) focuses on the band dispersion of DNL1 along $\Gamma$Z, and panels (k$_{1-5}$) correspond to perpendicular cuts along selected momenta $k_{\langle 001 \rangle}$ (white lines in \ref{fig4} j). Also here, we observe the evolution of a (remnant) Dirac crossing, in good agreement with the DFT results in panel (l). Our ARPES results thus clearly confirm the DNL scenario predicted by theory \cite{Sun2017,Ahn2019}.

\section{(110) Fermi surface along XRX.}

Surprises come upon inspection of a (110) plane containing the X and R high symmetry points, outlined in Fig.~\ref{fig5}~(a). Figure~\ref{fig5}~(b) shows an ARPES Fermi surface (b$_1$) and a constant energy cut at -80~meV (b$_2$), taken with $h\nu = 69$~eV photon energy, and probing the RuO$_2$ BZ along the (110) plane in Fig.~\ref{fig5}~(a) as discussed in section~\ref{sec: Navigating $k$-space}. We observe four main spectral contributions, marked in panel (b$_1$) and summarized in the schematics of panel (c). 

(I) First, and in direct contrast to Fig.~\ref{fig4}~(d), we now observe double arc structures centered at the X points. These represent the two branches of the Dirac crossing that form the unexpected DNL3 along XR, highlighted in (b$_2$). The arcs extend towards the zone center, and form a faint onion-dome. Fig.~\ref{fig5}~(d) shows horizontal band structure cuts for five selected momenta $k_{\langle 001\rangle}$ (black dashed lines in \ref{fig5}~c), revealing the evolution of the Dirac crossing from X towards R. From $k_{\langle 001\rangle} = 0$ (d$_1$) to $\sim0.25$~\AA$^{-1}$ (d$_5$), the crossing point moves towards lower binding energies (black arrows), and eventually passes the Fermi level at $k_{\langle 001\rangle} \sim 0.28$~\AA$^{-1}$. Our bulk DFT calculation in panel (e) reproduces the Dirac crossing in (d$_1$) astonishingly well, identifies its predominant $d_{x^2-y^2}$ character, but locates it $0.56$~eV above the experimental value of $\sim-0.1$~eV, a deficit of our simplified bulk DFT approach. The corresponding 3D band structure model of Fig.~\ref{fig5}~(f) correctly produces the continuous 4-fold band crossing of DNL3 along XR. This degeneracy however is lifted by SOC as seen in Fig.~\ref{fig5}~(g). As the degeneracy is strictly symmetry protected along the XM line (which also protects DNL2) \cite{Sun2017}, the SOC induced splitting effect is weak in the vicinity of the X point and remains unresolved by our ARPES experiment.

(II) Second, we find intense spectral features labeled DP in Fig.~\ref{fig5}~(b). These are the intersection points of DNL1 with the XRX momentum plane, as outlined in panel (a). Both ARPES and DFT reveal the corresponding Dirac crossing in panel~(h), but the SOC induced gap remains again unresolved, and bulk DFT locates the crossing point again about $0.15$~eV above the experimental value of -10~meV.

(III) Third, intense features dubbed `hot streaks' in Fig.~\ref{fig5}~(b) mark the projections of DNL1 onto the XRX plane (see Fig.~\ref{fig5}~a). Their intersections with the XR BZ boundary line mark the `hot spot' features in the Fermi surface, claimed responsible for an antiferromagnetic spin density wave instability in RuO$_2$ in Ref.~\onlinecite{Berlijn2017}. Fig.~\ref{fig5}~(i) shows ARPES cuts along $k_{\langle 001\rangle}$, taken at representative momenta $k_{\langle 110\rangle}$, as outlined in (c). Next to the features forming the onion-dome, we observe the continuous evolution of the Dirac states at DP in (i$_1$) towards a band with a hole-like parabolic band maximum at the `hot spots' in (i$_5$), as correctly predicted by bulk DFT (blue in panel j). This evolution is smooth and responsible for the intense `hot streaks' in the Fermi surface of panel (b), the projection of DNL1 onto the XRX momentum plane (see panel a). The strong nesting of these parallel `hot streaks' along commensurate nesting vectors, as well as their simultaneous electron- and hole-like character, might indeed favor potential Fermi surface instabilities such as spin- or charge density waves. In addition, the intersections of these `hot streaks' with the XR BZ boundary lines, i.e. the `hot spots', are symmetry protected by the non-symmorphic glide plane of RuO$_2$. The 4-fold degeneracy of these bands is thus lifted only by SOC (panel j), and/or by a Fermi surface instability \cite{Berlijn2017,Ahn2019}.

(IV) Last, we find two prominent arcs spanning in between adjacent DNL3s, the FBSS. The energy dispersion of the FBSS along $k_{\langle 110 \rangle}$, as well as its anchoring close to the Dirac crossing, is traced in Fig.~\ref{fig5}~(d). The bulk DFT description in panel (e) misses this state, which clearly demonstrates its surface character. Far away from X, the FBSS remains non-dispersively flat at $\sim -30$~meV, but takes a sharp, hole-like downward bend to merge with the Dirac crossing at the BZ boundary XR line. The ARPES cuts in Fig.~\ref{fig5}~(i) present the perpendicular dispersion of the FBSS at the BZ center (i$_1$), and trace its evolution with $k_{\langle 110 \rangle}$ (i$_{2-4}$) as it integrates into the DNL3 in (i$_5$). Along XR (i$_5$), DNL3 and the FBSS produce an electron-like parabolic dispersion (black dotted line), with a $\sim0.1$~eV band bottom and $m^*\sim2.5 m_e$ effective mass, well mimicked by the bulk DFT bands (red) in panel (j). The simultaneous electron- and hole character, as well as the diverging density of states of the FBSS, are clear hallmarks of a saddle-point van Hove singularity.

\figseven

\section{Doping evolution of the FBSS}\label{sec: doping evolution}

The spanning and anchoring of the FBSS in between adjacent DNL3s, as well as its flat energy dispersion, suggests this state to represent a topologically trivial analogue of the drumhead surface state predicted in systems with closed contour DNLs \cite{Weng2015,Chan2016}. We test its robustness by potassium deposition onto the surface while we monitor the ARPES response \textit{in situ}. An overview of the results is presented in Fig.~\ref{fig6}. Panels (a) and (b) show the continuous doping evolution of energy distribution curves (EDC) at $(k_{\langle 110\rangle},k_{\langle 001\rangle})=(0,0)$ and at the $X$ point, respectively. With increasing electron doping, the FBSS considerably broadens and disperses to $\sim-0.43$~eV (red dashed in Fig.~\ref{fig6}~a,b). The states associated with DNL3 populate only slightly (black dashed in Fig.~\ref{fig6}~b), producing the Fermi surface bifurcation in the momentum distribution curve (MDC) of panel (c). 

Panel (d) shows the K $1s$ core level peak, panels (e-h) show the ARPES data before and after potassium deposition, respectively. With respect to the pre-deposition Fermi surface in (e$_1$), the post-deposition Fermi surface (e$_2$) reveals overall broader and fuzzier spectral weight. However, whereas the bulk derived spectral contributions related to DNL3 (I), DP (II), and the `hot spot' states (III) remain intact, the FBSS disappears and gives way to the faint circular contours of a gas of itinerant surface electrons (2DEG). The dispersion of the FBSS along $k_{\langle 110\rangle}$ (white dashed line marked 'f' in e$_1$) before and after deposition is shown in panels (f$_1$) and (f$_2$), respectively. Dropping to higher binding energy, the FBSS produces broad but robust spectral weight at $\sim-0.43$~eV, while the 2DEG forms a broad parabolic line shape close to $E_F$. The drop of the FBSS is reproduced in panels (g) along $k_{\langle 001\rangle}$ (marked 'g' in e$_1$). The bulk band derived Dirac crossing at DP however stays remarkably intact. In panels (h), the `hot spot' states gain overall spectral weight with respect to DNL3, seemingly connect to the FBSS, and form a continuous M-shaped like band contour along $k_{\langle 001\rangle}$ (marked 'h' in e$_1$).

\figeight
\fignine

\section{ARPES versus DFT}\label{sec: ARPESvsDFT}

As pointed out in the previous sections, bulk DFT captures the overall ARPES band structure very well, but consistently misses the correct binding energy of some relevant bands by significant values. Figure~\ref{fig7} shows an overview of our bulk DFT calculations. Panels (a) and (b) again outline the $\Gamma$MAZ and XRX planes within the 3D BZ of RuO$_2$, respectively. The Kohn-Sham eigensolutions without and with SOC are shown along an augmented path in panel (c) and (d). The path covers the Z$\Gamma$MAZ as well as the corresponding projected path (ZA)($\Gamma$M)X(M$\Gamma$)(AZ)R(ZA) within the XRX plane, where `(xy)' denotes the midpoint in between high symmetry point 'x' and 'y'. Along these paths, the ARPES data are plotted for comparison. The calculation finds the octuple crossing OP at $\sim70$~meV, in acceptable agreement with our experiment. DFT however finds the hexatruple crossing point HP at $\sim0.66$~eV, i.e. 0.79~eV higher as the experimental value of -0.13~eV. The onset of DNL3 at X is predicted at $\sim0.46$, i.e. 0.56~eV higher than the ARPES finding at -105~meV. Last, the Dirac crossing point DP is predicted at $\sim0.14$~eV by DFT, i.e. 0.15~eV higher than the experimental finding of -10~meV. 

We resolve these discrepancies -- already raised in our previous work \cite{Jovic2018} -- by DFT calculations of a RuO$_2$ (110) surface slab. Figure~\ref{fig7_1}~(a) reproduces the ARPES data along RXR (a$_1$) and the perpendicular direction XX (a$_2$), showing the continuous Dirac crossing forming DNL3, the accompanying FBSS, as well as the `hot spot' states. In panels (b-d), we present the corresponding slab calculations for the thermodynamically favored oxygen rich- (b) as well as the stoichiometric- (c) and the ruthenium rich RuO$_2$ (110) surface (d) \cite{Over2012,Weaver2013}, plotted on top of the (110) surface projection of the bulk bands. The marker size of the DFT bands indicates their surface character. Apart from a rigid energy shift of $\sim 130$~meV (blue arrow) and a small renormalization factor, the calculation results for the oxygen rich surface in (b) match our ARPES results in (a) very well. In particular, theory captures the topologically trivial nature and the dispersion of the FBSS as well as its anchoring slightly below the DNL3, and identifies its predominant out of plane Ru $d_{z^2}$ orbital character. As the terminating ruthenium atoms on the RuO$_2$ surface form well separated chains along the $\langle 001 \rangle$ direction (see e.g. Fig.~27 in Ref.~\onlinecite{Over2012}), the band width of the FBSS is strongly suppressed along $\langle 110 \rangle$.

For the stoichiometric surface (c), without the top most layer of oxygen (O$_{\text{ot}}$) but only the bridging oxygen species (O$_{\text{br}}$) present (see Fig.~\ref{fig1}~c), the FBSS band-width is further decreased, while states along XR gain considerably in surface character. This trend continues for the ruthenium rich surface (d), which lacks both the O$_{\text{ot}}$ and O$_{\text{br}}$ species. The FBSS does not stabilize anymore, while states around the Dirac crossing at X as well as `hot spot' states gain considerably in surface character, and transfer spectral weight towards lower energies. This crossover from the oxygen rich to the ruthenium rich, i.e. oxygen poor surface, is remarkably similar to the crossover behavior of ARPES upon potassium deposition, discussed in section \ref{sec: doping evolution}, and reproduced in Fig.~\ref{fig7_1}~(e). It strongly suggests oxygen deficiencies at the (110) surface of RuO$_2$ to act as effective electron donors, similar to what is commonly observed in titanates such as TiO$_2$ anatase \cite{moser2013tunable}. It also suggests that the FBSS is involved in surface catalytic model reactions involving the adsorption of hydrogen (H$_2$), nitric oxide (NO) or carbon monoxide (CO). These gases are known to produce strong and reversible electron acceptor states, and are thus expected to populate or deplete the FBSS \cite{Wang2011a,Weaver2013,Over2012}.

\section{Discussion \& Outlook}

In summary, our APRES data confirms and expands earlier predictions of a DNL network in RuO$_2$ by Sun et al. \cite{Sun2017}. We find an additional DNL3 close to the Fermi level, that escaped previous DFT investigations \cite{Sun2017}. This DNL3 remains remarkably robust despite considerable SOC, whose effect we find to be rather weak in proximity of the X-point, and beyond the resolution of our experiment. This bears some similarity to graphene, a predicted quantum spin Hall insulator \cite{Kane2005}, which in view of weak SOC due to small next-nearest neighbor hybridization presents itself -- like RuO$_2$ -- as a de facto Dirac semi-metal \cite{Bostwick2007}. Our results constitute a direct observation of Dirac physics at the Fermi level for a functional oxide of genuine industrial importance. 

We further wish to emphasize the similarity between the topologically trivial FBSS in RuO$_2$ (110) and FBSSs that are predicted to span in between the surface projections of topologically non-trivial DNLs. Given the particular surface structure of RuO$_2$ (110), we interpret this aspect to date as merely coincidental. It is worth noting, however, that the saddle point van Hove singularity shaped dispersion of the FBSS locates close to the Fermi level. For such a van Hove singularity, Coulomb interaction and correlation effects of the surface electrons might be considerably enhanced \cite{Huh2016}. Thus, even in the presence of relatively weak perturbations, we may expect exotic symmetry broken ground states such as surface magnetism \cite{Magda2014,Chan2016}, surface superconductivity \cite{Kopnin2011,Tang2014}, or graphene-like Landau levels \cite{Rhim2015}. Several aspects lend themselves for future investigation, i.e., the tunability of the FBSS by well studied surface reactions, such as the adsorption of hydrogen (H$_2$), nitric oxide (NO), or carbon monoxide (CO) \cite{Wang2011a,Weaver2013,Over2012}. We expect such reactions to $p$-dope the van Hove singularity towards the Fermi level, which is again a line of investigation left for future sutdies.

Finally, already for pristine RuO$_2$ as studied here, we note the similarity of our ARPES data to recent theorectical findings predicting antiferromagnetic Fermi surface instabilities of spin-density wave- \cite{Berlijn2017,Zhu2018} and Pomeranchuk-type \cite{Ahn2019}, which in principle could be accessible by spin resolved ARPES experiments. While magnetic moments in general may have a great impact on the catalytic efficiency \cite{Torun2013}, the collinear antiferromagnetic ordering of RuO$_2$ may break the glide plane symmetries that prevent the system from a non-zero anomalous Hall conductance, and lead to novel phenomena such as the recently predicted crystal Hall effect \cite{Smejkal2019}.

\begin{acknowledgments}
We thank Johan Chang, Masafumi Horio, Paul Snijders and Yan Sun for helpful discussions. S.M. was supported by the Swiss National Science Foundation (Grant No. P300P2-171221). R.J.K. was supported by a fellowship within the Postdoc-Program of the German Academic Exchange Service (DAAD). D.D.S. and R.T. acknowledge the DFG through SFB1170 ``Tocotronics'' and the ERC-StG-336012-Thomale-TOPOLECTRICS, as well as the Gauss Centre for Supercomputing e.V. (www.gauss-centre.eu) for providing computing time on the GCS Supercomputer Super-MUC at Leibniz Supercomputing Centre (www.lrz.de). We further acknowledge financial support from the DFG through the W\"urzburg-Dresden Cluster of Excellence on Complexity and Topology in Quantum Matter -- \textit{ct.qmat} (EXC 2147, project-id 39085490). The Boston University program was supported by the Department of Energy under Grant No. DE-FG02-98ER45680. The bulk DFT work was supported by a Consolidator grant of the European Research Council under project number 617196, and used resources of IDRIS/GENCI under project gen1393. We thank the computer support team of CPHT. This research further used resources of the Advanced Light Source, which is a DOE Office of Science User Facility under contract no. DE-AC02-05CH11231.
\end{acknowledgments}

\nocite{*}


%

\end{document}